\documentclass[prd,superscriptaddress,amsfonts,amssymb,amsmath,showpacs,twocolumn]{revtex4-2}
\usepackage{bm}
\usepackage{amsfonts}
\usepackage{latexsym}
\usepackage[latin1]{inputenc}
\usepackage{graphicx}
\usepackage{amsmath}
\usepackage{palatino}
\usepackage{ragged2e}
\usepackage{mathpazo}
\usepackage{textcomp}
\usepackage{float}
\usepackage{booktabs}
\usepackage{dcolumn}
\usepackage{multirow}
\usepackage{hyperref}
\hypersetup{colorlinks,citecolor=blue}
\usepackage{amsmath}
\usepackage{xcolor}
\usepackage{orcidlink}
\usepackage[caption=false]{subfig}
\usepackage{commath}
\def\jnl@style{\it}
\def\aaref@jnl#1{{\jnl@style#1}}

\def\aaref@jnl#1{{\jnl@style#1}}

\def\aj{\aaref@jnl{AJ}}                   
\def\apj{\aaref@jnl{ApJ}}                 
\def\apjl{\aaref@jnl{ApJ}}                
\def\apjs{\aaref@jnl{ApJS}}               
\def\apss{\aaref@jnl{Ap\&SS}}             
\def\aap{\aaref@jnl{A\&A}}                
\def\aapr{\aaref@jnl{A\&A~Rev.}}          
\def\aaps{\aaref@jnl{A\&AS}}              
\def\mnras{\aaref@jnl{Mon.~Not.~Roy.~Astron.~Soc.}}             
\def\prd{\aaref@jnl{Phys.~Rev.~D}}        
\def\prc{\aaref@jnl{Phys.~Rev.~C}}  
\def\prl{\aaref@jnl{Phys.~Rev.~Lett.}}    
\def\qjras{\aaref@jnl{QJRAS}}             
\def\skytel{\aaref@jnl{S\&T}}             
\def\ssr{\aaref@jnl{Space~Sci.~Rev.}}     
\def\zap{\aaref@jnl{ZAp}}                 
\def\nat{\aaref@jnl{Nature}}              
\def\aplett{\aaref@jnl{Astrophys.~Lett.}} 
\def\apspr{\aaref@jnl{Astrophys.~Space~Phys.~Res.}} 
\def\physrep{\aaref@jnl{Phys.~Rep.}}      
\def\physscr{\aaref@jnl{Phys.~Scr}}       
\def\commat{\aaref@jnl{Comm.~Math.~Phys.}}              
\def\science{\aaref@jnl{Science}}               
\def\cqg{\aaref@jnl{Classical Quant.~Grav.}}            
\def\jpcs{\aaref@jnl{JPCS}}                                     
\def\ijmpd{\aaref@jnl{Int.~J.~Mod.~Phys.~D}}                    
\def\grg{\aaref@jnl{Gen.~Relat.~Gravit.}}               
\def\rpp{\aaref@jnl{Rep.~Prog.~Phys.}}          
\def\npa{\aaref@jnl{Nucl.~Phys.~A}}        
\def\lrr{\aaref@jnl{Living Rev.~Rel.}}                   
\def\jcap{\aaref@jnl{J.~Cosmology Astropart.~Phys.}}    
\def\rmp{\aaref@jnl{Rev.~Mod.~Phys.}}   
\def\epjc{\aaref@jnl{Eur.~Phys.~J.~C}}

\allowdisplaybreaks[1]

\begin{document}

\color{black}       

\title{Investigating the dark energy phenomenon in $f(R,L_m)$ cosmological models with observational constraints}

\author{S. Myrzakulova\orcidlink{0000-0002-0027-0970}}
\email[Email: ]{shamyrzakulova@gmail.com}
\affiliation{L. N. Gumilyov Eurasian National University, Astana 010008,
Kazakhstan.}
\affiliation{Ratbay Myrzakulov Eurasian International Centre for Theoretical
Physics, Astana 010009, Kazakhstan.}

\author{M. Koussour\orcidlink{0000-0002-4188-0572}}
\email[Email: ]{pr.mouhssine@gmail.com}
\affiliation{Quantum Physics and Magnetism Team, LPMC, Faculty of Science Ben
M'sik,\\
Casablanca Hassan II University,
Morocco.}

\author{N. Myrzakulov\orcidlink{0000-0001-8691-9939}}
\email[Email: ]{nmyrzakulov@gmail.com}
\affiliation{L. N. Gumilyov Eurasian National University, Astana 010008,
Kazakhstan.}
\affiliation{Ratbay Myrzakulov Eurasian International Centre for Theoretical
Physics, Astana 010009, Kazakhstan.}

\date{\today}
\begin{abstract}

This paper explores the dark energy phenomenon within the context of $f(R,L_m)$ gravity theory. Two specific non-linear $f (R, L_m)$ models are considered: $f(R,L_m)=\frac{R}{2}+L_m^\alpha$ and $f(R,L_m)=\frac{R}{2}+(1+\alpha R)L_m$, where the parameter $\alpha$ is free. Here, we adopt a parametrization form for the Hubble parameter in terms of redshift $z$ as $H(z)=H_0 \left[A(1+z)^3+B+\epsilon \log(1+z)\right]^\frac{1}{2}$, which allows for deviations from the standard $\Lambda$CDM model at both low and high redshifts. We then incorporate the Hubble parameter solution into the Friedmann equations for both models. We employ Bayesian analysis to estimate the
constraints on the free parameters $H_0$, $A$, $B$, and $\epsilon$ using the Hubble measurements and the Pantheon dataset. Further, we investigate the evolution of key cosmological quantities, such as the deceleration parameter, energy density, pressure, EoS parameter, and energy conditions. The evolution of the deceleration parameter reveals a significant transition from a decelerating phase to an accelerating phase in the Universe. The EoS parameter exhibits quintessence-like behavior for both non-linear $f (R, L_m)$ models.

\textbf{Keywords:} Cosmology, Dark energy, $f(R,L_m)$ gravity, Observational constraints, Energy conditions.

\end{abstract}

\maketitle

\section{Introduction}\label{sec1}

We maintain our belief that Einstein's theory of General Relativity (GR) does not provide the ultimate explanation for all gravitational phenomena, despite its successful validation in solar system tests \cite{Will}. Through extensive observations and experiments, it has become evident that certain phenomena, such as the cosmic accelerating expansion and the mysteries surrounding Dark Matter (DM), suggest the need for a potential modification of GR at large scales. The discovery of the Universe's expansion accelerating in its later stages \cite{Riess,Perlmutter,C.L.,R.R.,D.N.,D.J.,W.J.,T.Koivisto,S.F.} has served as a direct motivation for pursuing such modifications. The simplest approach to account for the observed acceleration of the Universe is to introduce a cosmological constant, denoted as $\Lambda$, which effectively acts as Dark Energy (DE), which fills the Universe and accounts for around 70\% of its total energy content. This DE component is responsible for generating a negative pressure that drives the accelerated expansion. The $\Lambda$ Cold DM ($\Lambda$CDM) model faces certain challenges, such as the coincidence problem \cite{Dalal}, which refers to the fact that the densities of non-relativistic matter and DE are of the same order at present. Another intricate issue associated with the cosmological constant is the cosmological constant problem, which involves the significant inconsistency between the observed value of $\Lambda$ based on astronomical observations \cite{Riess,Perlmutter} and the theoretically predicted value of the quantum vacuum energy from particle physics \cite{S.W.}. In order to address the aforementioned cosmological challenges, various dynamical (time-varying) DE models have been proposed in the literature \cite{Phan1,Phan2,Phan3,Quin}. 

Alternatively, DE can also be understood as an effective geometric quantity arising from modifications to the Einstein-Hilbert (EH) action. In order to incorporate these modifications, we can replace the Ricci curvature, represented by $R$, in the EH with a generic function $f(R)$. This formulation gives rise to the class of theories known as $f(R)$ theories, which have been extensively studied and discussed in the literature \cite{H.A.,R.K.,H.K.}. Numerous investigations have been conducted on modified theories of gravity that can account for both the early and late-time expansion of the Universe. One notable example is the $f(Q)$ gravity, where the Ricci scalar $R$ is replaced by a general function $f(Q)$, with $Q$ representing the non-metricity scalar \cite{Q0,Q1}. Other modified theories of gravity, such as $f(T)$ where $T$ represents torsion \cite{T1, T2, T3, T4, T5}, $f(R, G)$, $f(R, T)$, and more, provide alternative explanations for the current accelerated expansion of the Universe without invoking an effective DE term, akin to a cosmological constant. Previous studies on these theories are outlined in the references provided \cite{Carr, Cap, LAM,Q2,Q3,Q4,Q5,Q6,Q7,Q8,Harko2011}.

Recently, Bertolami et al. \cite{RL0} introduced a generalized version of modified gravity, specifically the $f(R)$ theory, by incorporating a direct coupling between the generic $f(R)$ function, which describes the Ricci curvature scalar $R$, and the Lagrangian density of matter $L_m$ within EH action. Building upon this, Harko and Lobo extended the model to include arbitrary geometry-matter couplings \cite{THK}. The inclusion of non-minimal curvature and matter couplings in cosmological models has led to intriguing applications in cosmology and astrophysics \cite{THK-2,THK-3,V.F.-2}. Additionally, Harko and Lobo proposed the $f(R, L_m)$ gravity, which encompasses a broad range of curvature-matter coupling theories \cite{RL}. In this modified gravity framework, the energy-momentum tensor exhibits a non-vanishing covariant divergence, leading to the emergence of an additional force orthogonal to four velocities. Consequently, the motion of test particles deviates from geodesic paths. It is worth noting that the $f(R, Lm)$ modified theory of gravity does not adhere to the equivalence principle and is subject to constraints imposed by solar system tests \cite{FR,JP}. In recent years, there has been a surge of interest in the cosmological implications of the $f(R, Lm)$ gravity theory, resulting in a growing body of literature on the topic \cite{RL1,RL2,RL3,RL4}. 
A recent study by Wang and Liao \cite{WG} focused on investigating the energy conditions within the context of $f(R,L_m)$ gravity. Their work shed light on the implications of this modified gravity theory on energy conditions. In addition, Goncalves and Moraes \cite{GM} conducted an analysis of cosmological aspects by incorporating the non-minimal matter-geometry coupling in $f(R,L_m)$ gravity. Their study delved into the effects of this coupling on various cosmological phenomena.

Although numerous theoretical approaches have been proposed to explain the DE phenomenon, no definitive model has been identified as the correct one. One of the current approaches used to describe late-time cosmic acceleration is parametrization \cite{Pacif1,Pacif2}. Parametrization involves introducing specific mathematical functions or parameters, such as the deceleration parameter, Hubble parameter, and jerk parameter, to characterize the behavior of cosmic expansion. These parameters are chosen to effectively capture the observed accelerated expansion of the Universe. This approach allows for flexibility in modeling and provides a framework for studying the dynamics and properties of the accelerating Universe. It has been extensively studied in the literature and has been the subject of numerous investigations \cite{H1,H2,DP1,DP2,DP3}.

In this study, we adopt a parametrization form for the Hubble parameter in terms of redshift z, given by $H(z)=H_0 \left[A(1+z)^3+B+\epsilon \log(1+z)\right]^\frac{1}{2}$, where $H_0$, $A$, $B$, and $\epsilon$ are free parameters. This parametrization allows for deviations from the standard $\Lambda$CDM model at both low and high redshifts. We further investigate the FLRW Universe within the framework of $f(R,L_m)$ gravity by considering two non-linear $f(R,L_m)$ models, specifically, $f(R,L_m)=\frac{R}{2}+L_m^\alpha$ (model 1) and $f(R,L_m)=\frac{R}{2}+(1+\alpha R)L_m$ (model 2), where $\alpha$ represents a free model parameter. This paper is organized as follows: In Sec. \ref{sec2}, we present the formalism of $f(R,L_m)$ gravity in Flat FLRW Universe. In Sec. \ref{sec3}, we adopt a parametrization form to describe the relationship between the Hubble parameter and redshift z. Subsequently, we employ Bayesian analysis to estimate the constraints on the free parameters $H_{0}$, $A$, and $\epsilon$ using the Hubble measurements and the Pantheon dataset. We also examine the evolution of the deceleration parameter for the model parameters obtained from these different datasets. In Sec. \ref{sec4}, we focus on a specific $f(R, L_m)$ model, specifically $f(R,L_m)=\frac{R}{2}+L_m^\alpha$, where $\alpha$ is a free model parameter. We investigate the evolution of key cosmological quantities, such as energy density, pressure, EoS parameters, and energy conditions, in order to gain insights into the DE phenomenon. In addition, we explore another non-linear $f(R, L_m)$ model, namely $f(R,L_m)=\frac{R}{2}+(1+\alpha R)L_m$, and analyze the behavior of various cosmological parameters within this framework. Finally, in Sec. \ref{sec5}, we summarize and present our conclusions.
 
\section{Flat FLRW Universe in $f(R,L_m)$ Cosmology: A Fundamental Formulation}\label{sec2}

Here, we consider the expression for the action of $f(R,L_m)$ gravity theory as \cite{RL}
\begin{equation}\label{1}
S= \int{f(R,L_m)\sqrt{-g}d^4x}, 
\end{equation}
where $f(R,L_m)$ is an arbitrary function of the Ricci scalar $R$ and the matter Lagrangian $L_m$, $g$ represents the determinant of the metric tensor.

The Ricci scalar $R$ can be calculated by contracting the Ricci tensor $R_{\mu\nu}$ as 
\begin{equation}\label{2}
R= g^{\mu\nu} R_{\mu\nu},
\end{equation} 
where
\begin{equation}\label{3}
R_{\mu\nu}= \partial_\lambda \Gamma^\lambda_{\mu\nu} - \partial_\mu \Gamma^\lambda_{\lambda\nu} + \Gamma^\lambda_{\mu\nu} \Gamma^\sigma_{\sigma\lambda} - \Gamma^\lambda_{\nu\sigma} \Gamma^\sigma_{\mu\lambda},
\end{equation}
with $\Gamma^\alpha_{\beta\gamma}$ denotes the components of the Levi-Civita connection and can be derived as
\begin{equation}\label{4}
\Gamma^\alpha_{\beta\gamma}= \frac{1}{2} g^{\alpha\lambda} \left( \frac{\partial g_{\gamma\lambda}}{\partial x^\beta} + \frac{\partial g_{\lambda\beta}}{\partial x^\gamma} - \frac{\partial g_{\beta\gamma}}{\partial x^\lambda} \right)
\end{equation}

The gravitational field equation derived by taking the variation of the action \eqref{1} with respect to the metric tensor $g_{\mu\nu}$ is,
\begin{equation}\label{5}
f_R R_{\mu\nu} + (g_{\mu\nu} \square - \nabla_\mu \nabla_\nu)f_R - \frac{1}{2} (f-f_{L_m}L_m)g_{\mu\nu} = \frac{1}{2} f_{L_m} T_{\mu\nu}.
\end{equation}

Here, we introduce the notations $f_R \equiv \frac{\partial f}{\partial R}$ and $f_{L_m} \equiv \frac{\partial f}{\partial L_m}$. Furthermore, $T_{\mu\nu}$ represents the energy-momentum tensor for a perfect fluid, which is defined as
\begin{equation}\label{6}
T_{\mu\nu} = \frac{-2}{\sqrt{-g}} \frac{\delta(\sqrt{-g}L_m)}{\delta g^{\mu\nu}}.
\end{equation}

Furthermore, contracting the field equation \eqref{5} leads to the following relationship between the trace of the energy-momentum tensor $T$, the Lagrangian term $L_m$, and the Ricci scalar $R$,
\eqref{5},
\begin{equation}\label{7}
R f_R + 3\square f_R - 2(f-f_{L_m}L_m) = \frac{1}{2} f_{L_m} T,
\end{equation}
where $\square F$ denotes the d'Alembertian operator applied to a scalar function $F$, defined as $\square F = \frac{1}{\sqrt{-g}} \partial_\alpha (\sqrt{-g} g^{\alpha\beta} \partial_\beta F)$.

Moreover, upon taking the covariant derivative of Eq. \eqref{5}, we obtain the following result,
\begin{equation}\label{8}
\nabla^\mu T_{\mu\nu} = 2\nabla^\mu ln(f_{L_m}) \frac{\partial L_m}{\partial g^{\mu\nu}}.
\end{equation}

Now, by embracing the cosmological principle, we elucidate our understanding of the Universe by employing the spatially isotropic and homogeneous flat FLRW (Friedmann-Lema\^{i}tre-Robertson-Walker) metric. This metric serves as a fundamental tool in characterizing the overall structure and dynamics of our expansive cosmos. By assuming this metric, we can investigate the behavior of various cosmological phenomena, such as the evolution of the Universe, the distribution of matter, and the expansion rate, while considering the underlying principles of isotropy and homogeneity:
\begin{equation}\label{9}
ds^2= -dt^2 + a^2(t)[dx^2+dy^2+dz^2],
\end{equation}
where $a(t)$ denotes the scale factor that quantifies the cosmic expansion at a specific time $t$. By taking into account the line element (\ref{9}), we have calculated the Ricci scalar as
\begin{equation}\label{12}
R= 6 ( \dot{H}+2H^2 ),
\end{equation}
where $H=\frac{\dot{a}}{a}$ represents the Hubble parameter, which quantifies the rate of cosmic expansion at a particular epoch.

The energy-momentum tensor that describes the matter content of the Universe, characterized by isotropic pressure $p$ and energy density $\rho$, corresponding to the line element \eqref{9}, is given by
\begin{equation}\label{13}
\mathcal{T}_{\mu\nu}=(\rho+p)u_\mu u_\nu + pg_{\mu\nu}.
\end{equation}

Here, $u^\mu=(1,0,0,0)$ represents the four-velocity components that characterize the perfect cosmic fluid. 

The Friedmann equations, governing the dynamics of the Universe for the function $f(R,L_m)$, can be expressed as
\begin{equation}\label{14}
3H^2 f_R + \frac{1}{2} \left( f-f_R R-f_{L_m}L_m \right) + 3H \dot{f_R}= \frac{1}{2}f_{L_m} \rho,
\end{equation}
and
\begin{equation}\label{15}
\dot{H}f_R + 3H^2 f_R - \ddot{f_R} -3H\dot{f_R} + \frac{1}{2} \left( f_{L_m}L_m - f \right) = \frac{1}{2} f_{L_m}p.
\end{equation}  

\section{Energy Conditions}\label{sec3}

The energy conditions play a crucial role in understanding the geodesics of the Universe. Also, these conditions are relationships imposed on the energy-momentum tensor to ensure the presence of positive energy within the system. Derived from the well-known Raychaudhury equations \cite%
{Raychaudhuri, Nojiri2, Ehlers}, they are expressed as
\begin{equation}
\label{R1}
\frac{d\theta}{d\tau}=-\frac{1}{3}\theta^2-\sigma_{\mu\nu}\sigma^{\mu\nu}+\omega_{\mu\nu}\omega^{\mu\nu}-R_{\mu\nu}u^{\mu}u^{\nu}\,,
\end{equation}
\begin{equation}
\label{R2}
\frac{d\theta}{d\tau}=-\frac{1}{2}\theta^2-\sigma_{\mu\nu}\sigma^{\mu\nu}+\omega_{\mu\nu}\omega^{\mu\nu}-R_{\mu\nu}n^{\mu}n^{\nu}\,,
\end{equation}
where $\theta$ denotes the expansion factor, $n^{\mu}$ represents the null vector, and $\sigma^{\mu\nu}$ and $\omega_{\mu\nu}$ are the shear and rotation associated with the vector field $u^{\mu}$. In the case of attractive gravity, Eqs. \eqref{R1} and \eqref{R2} satisfy the following conditions:
\begin{align}
R_{\mu\nu}u^{\mu}u^{\nu}\geq0\,,\\
 R_{\mu\nu}n^{\mu}n^{\nu}\geq0\,.
\end{align}

Hence, when dealing with a perfect fluid matter distribution in the context of modified gravity, the energy conditions can be stated as 
\begin{itemize}
\item Null energy condition (NEC): $\rho_{eff}+p_{eff}\geq 0\,$;

\item Weak energy conditions (WEC): $\rho_{eff}\geq 0$ and $\rho_{eff}+p_{eff}\geq 0\,$;

\item Dominant energy conditions (DEC): $\rho_{eff}\geq 0$ and $|p_{eff}|\leq \rho_{eff}\,$.

\item Strong energy conditions (SEC):  $\rho_{eff}+3p_{eff}\geq 0\,$;
\end{itemize}
where $\rho_{eff}$ and $p_{eff}$ represent the effective energy density and effective pressure, respectively.

\section{Analyzing Data Using $H(z)$ Parameterization}\label{sec4}

To comprehend the cosmos and analyze its evolution, various approaches can be employed, including the parameterization of the Hubble parameter. By establishing a relationship between the Hubble parameter $H$ and the cosmic redshift parameter $z$, we can explore the mathematical expressions that describe the evolution of the cosmological models under investigation. This approach entails incorporating corrections associated with the DE component, extending beyond the standard $\Lambda$CDM model that includes the cosmological constant and CDM. In the standard $\Lambda$CDM cosmology, the Hubble parameter is expressed as
\begin{equation}
   H(z)=H_0 \left[\Omega_{m0}(1+z)^3+\Omega_{\Lambda}\right]^\frac{1}2.
   \label{LCDM}
\end{equation}

Here, $H_0$, $\Omega_{m0}$ and $\Omega_{\Lambda}=(1-\Omega_{m0})$ represent the present values of the Hubble parameter, total matter density, and total DE density, respectively. While our current comprehension of cosmic evolution is primarily founded on the $\Lambda$CDM model, it is important to acknowledge that there are notable discrepancies between the predictions of this theory and the observed phenomena. These deviations have prompted extensive investigations and discussions on the cosmological challenges, which are covered in detail in the references cited in the introduction \cite{Dalal,S.W.}. Therefore, we adopt a parameterized cosmological model that allows deviations from the standard $\Lambda$CDM model at both low and high redshifts. Specifically, we consider a logarithmic correction associated with the DE term in the following parametric form for $H(z)$,
\begin{equation}
   H(z)=H_0 \left[A(1+z)^3+B+\epsilon \log(1+z)\right]^\frac{1}{2},
   \label{Hz}
\end{equation}
where $H_0$, $A$, $B$, and $\epsilon$ are free parameters. In our model, $A$ is selected to represent the matter density parameter at the present epoch ($A=\Omega_{m0}$), whereas $\epsilon$ introduces a logarithmic term to potentially capture certain behaviors that are not accounted for in simpler parameterizations such as the $\Lambda$CDM model. In addition, to obtain $H=H_0$ at $z=0$, it is necessary for the variable $B=(1-A)$. The parameterization of Eq. (\ref{LCDM}) can be reproduced if the following condition is satisfied: $A=\Omega_{m0}$, $B=(1-\Omega_{m0})$, and $\epsilon=0$. On the other hand, the term $\epsilon \log(1+z)$ introduces a logarithmic correction, which is motivated by our intention to create a more flexible model capable of capturing potentially more complex behaviors in the expansion rate of the Universe. This choice is driven by the desire to explore deviations from the standard $\Lambda$CDM model, which assumes a simple, constant cosmological constant ($\Lambda$) and a matter density parameter. The inclusion of the log term introduces an additional degree of freedom to the model, enabling it to account for variations in the expansion rate that simpler parameterizations may overlook. This increased flexibility is valuable for assessing the model's compatibility with observational data and determining whether it can offer a better description of the observed behavior of the universe \cite{Log1,Log2}.

To investigate the behavior of cosmological parameters with respect to redshift, it is essential to consider the relationship between redshift and the Universe's scale factor: $a(t) = 1/(1+z)$. Thus, we can describe the correlation between the derivative of the Hubble parameter with respect to time and redshift as
\begin{equation}
\dot{H}=\frac{dH}{dt}=-(1+z)H(z)\frac{dH}{dz}
\end{equation}

The dynamics and essential cosmological characteristics of the model described in Eq. \eqref{Hz} are intricately tied to the model parameters ($H_0$, $A$, $\epsilon$). In the following subsection, we apply constraints to the model parameters ($H_0$, $A$, $\epsilon$) using up-to-date observational datasets.

Now, to constrain the parameters $H_0$, $A$, and $\epsilon$, one can leverage various observational datasets. In this study, we employ the standard Bayesian technique and utilize the Markov Chain Monte Carlo (MCMC) method to obtain the posterior distributions of the parameters \cite{BS}. The MCMC analysis is conducted using the \texttt{emcee} package \cite{Mackey/2013}. Specifically, we incorporate two datasets: the Hubble measurements (referred to as $Hz$ data) and the Pantheon dataset (referred to as $SNe$ data). The likelihood function defined below is employed to determine the best-fit values of the parameters:
\begin{equation}
    \mathcal{L} \propto exp(-\chi^2/2),
\end{equation}
where $\chi^2$ is the chi-squared function \cite{BS}. In the following, we present the expressions for the $\chi^2$ functions utilized for different datasets:

\subsection{$Hz$ dataset}

Singirikonda and Desai \cite{Singirikonda} recently compiled a collection of Hubble measurements within the redshift range of $0.07<z<1.965$. This dataset, referred to as the $Hz$ dataset, was obtained from the differential ages (DA) $\Delta t$ of galaxies \cite{D1,D2,D3,D4}. The comprehensive list of datasets can be found in \cite{Moresco/2018}. In order to estimate the model parameters $H_0$, $A$, and $\epsilon$, we employ the chi-square function as
\begin{equation}\label{25}
\chi _{Hz}^{2}(H_0,A,\epsilon)=\sum\limits_{k=1}^{31}
\frac{[H_{th}(z_{k},H_0,A,\epsilon)-H_{obs}(z_{k})]^{2}}{
\sigma _{H(z_{k})}^{2}},  
\end{equation}
where $H_{th}$ denotes the theoretical value of the Hubble parameter obtained from our model, while $H_{obs}$ represents its observed value. The uncertainty related to the Hubble parameter measurements is captured by the standard deviation $\sigma_{H(z_{k})}$.

\begin{figure}[h]
\centerline{\includegraphics[scale=0.65]{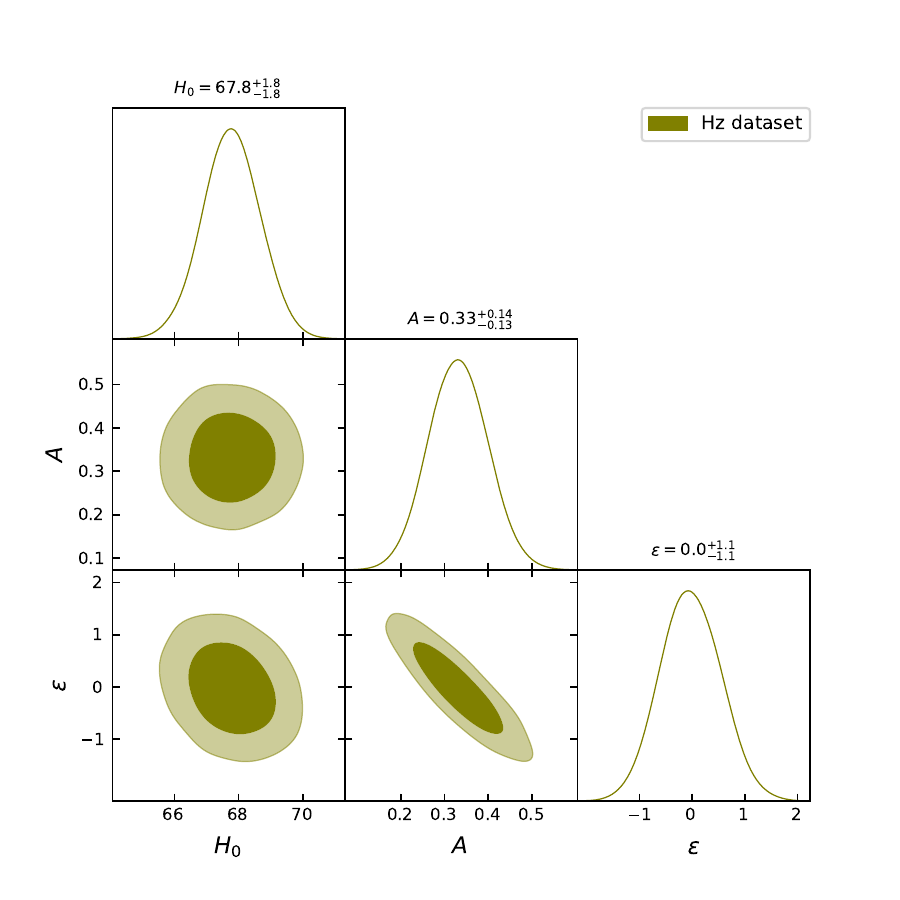}}
\caption{Likelihood contours for model parameters using the $Hz$ dataset: $1-\sigma$ and $2-\sigma$ confidence intervals depicted.}
\label{F_Hz}
\end{figure}

Fig. \ref{F_Hz} presents a contour plot illustrating the model parameters $H_{0}$, $A$, and $\epsilon$ based on the $Hz$ dataset. The contours depict the $1-\sigma$ and $2-\sigma$ confidence intervals. The obtained best-fit values for the model parameters are as follows: $H_{0}=67.8_{-1.8}^{+1.8}$ $km/s/Mpc$, $A=0.33_{-0.13}^{+0.14}$, and $\epsilon=0.0_{-1.1}^{+1.1}$. In this case, the observational constraints on $H(z)$ are consistent with the predictions of the $\Lambda$CDM model, where $\epsilon=0$. The value of $H_0$ obtained in the analysis, which is close to the value measured by recent Planck measurements for the base-$\Lambda$CDM cosmology, reinforces the consistency of the findings. The measured value of $H_0=67.4 \pm{0.5}$ $km/s/Mpc$ from the Planck mission is widely regarded as one of the most precise measurements of the Hubble constant \cite{planck_collaboration/2020}.

\subsection{$SNe$ dataset}

In this study, we employ the up-to-date Pantheon $SNe$ Ia dataset, which comprises 1048 data points obtained from various surveys such as SNLS, SDSS, Pan-STARRS1, HST, and low-redshift observations. This sample covers the redshift range of $z \in [0.01,2.3]$ and is utilized to constrain the aforementioned parameters \cite{Scolnic/2018}. The chi-squared function $\chi^2_{SNe}$ is derived from the Pantheon sample consisting of 1048 $SNe$ Ia data points \cite{Scolnic/2018} and can be expressed as
\begin{equation}\label{27}
\chi^2_{SNe}(H_0,A,\epsilon)=\sum_{i,j=1}^{1048}\Delta\mu_{i}\left(C^{-1}_{SNe}\right)_{ij}\Delta\mu_{j},
\end{equation}
where $C_{SNe}$ denotes the covariance metric \cite{Scolnic/2018}, and
\begin{equation}
  \quad \Delta\mu_{i}=\mu^{th}(z_i,H_0,A,\epsilon)-\mu_i^{obs}.
\end{equation}

In this context, $\mu^{th}$ represents the theoretical value of the distance modulus, while $\mu^{obs}$ corresponds to its observed value. The distance modulus is computed theoretically as
\begin{equation}\label{28}
\mu^{th}(z)= 5log_{10}D_{L}(z)+\mu_{0}, 
\end{equation}
with 
\begin{equation}\label{29}
\mu_{0} =  5log(1/H_{0}Mpc) + 25.
\end{equation}

The luminosity distance $D_L$ can be computed from the Hubble parameter as \cite{planck_collaboration/2020}
\begin{equation}\label{26}
D_{L}(z)= c(1+z) \int_{0}^{z} \frac{ dz'}{H(z')},
\end{equation}
where $c$ is the speed of light.

\begin{figure}[h]
\centerline{\includegraphics[scale=0.65]{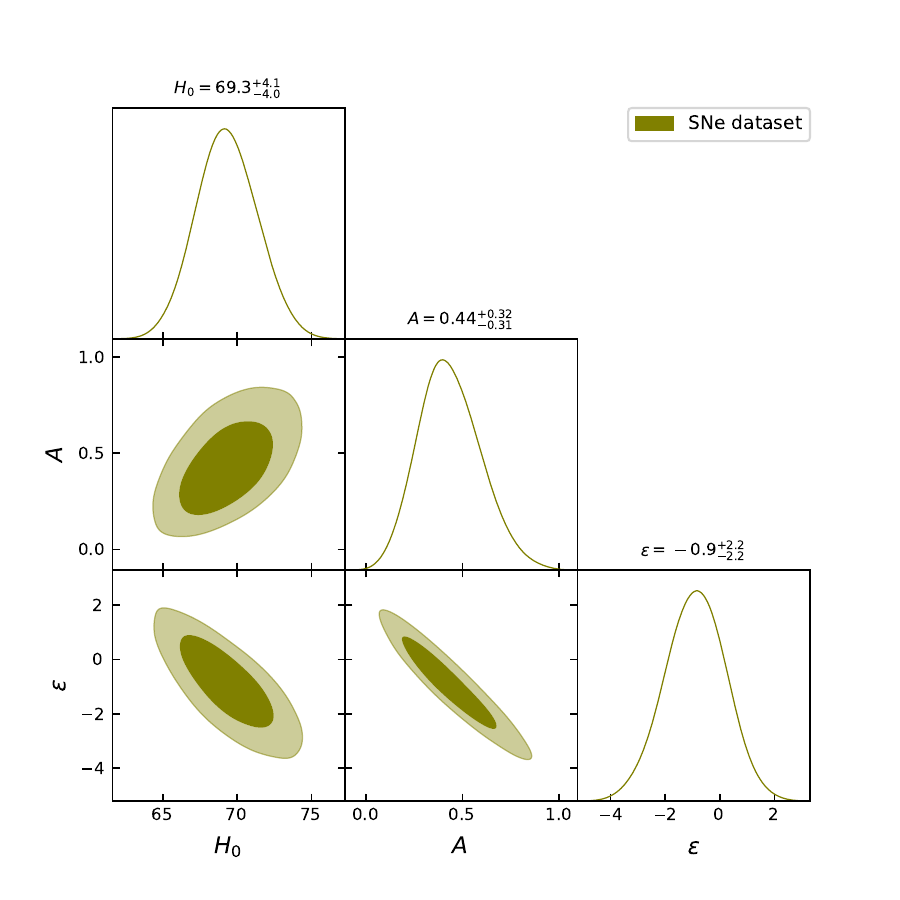}}
\caption{Likelihood contours for model parameters using the $SNe$ dataset: $1-\sigma$ and $2-\sigma$ confidence intervals depicted.}
\label{F_SNe}
\end{figure}

Fig. \ref{F_SNe} presents a contour plot illustrating the model parameters $H_{0}$, $A$, and $\epsilon$ based on the $SNe$ dataset. The contours depict the $1-\sigma$ and $2-\sigma$ confidence intervals. The obtained best-fit values for the model parameters are as follows: $H_{0}=69.3_{-4.0}^{+4.1}$ $km/s/Mpc$, $A=0.44_{-0.31}^{+0.32}$, and $\epsilon=-0.9_{-2.2}^{+2.2}$. Recently, Freedman et al. \cite{Freedman} conducted an independent measurement of the Hubble constant using the tip of the red giant branch as a distance estimator. Their analysis yielded a value of $H_0 = 69.8 \pm 1.9$ $km/s/Mpc$, which is in remarkable agreement with the value obtained in our model.

\subsection{$Hz+SNe$ dataset}

To obtain comprehensive constraints on the parameters $H_{0}$, $A$, and $\epsilon$ from the $Hz$ and $SNe$ datasets, we employ the total likelihood function. This joint likelihood function is obtained by taking the product of the likelihood functions for the $Hz$ data and $SNe$ data,
\begin{equation}
    \mathcal{L}_{Joint}= \mathcal{L}_{Hz} \times \mathcal{L}_{SNe}.
\end{equation}

Likewise, the joint chi-square function is obtained by summing the individual chi-square functions for the $Hz$ data and $SNe$ data,
\begin{equation}
    \chi^{2}_{Joint} = \chi^{2}_{Hz} + \chi^{2}_{SNe}.
\end{equation}

\begin{figure}[h]
\centerline{\includegraphics[scale=0.65]{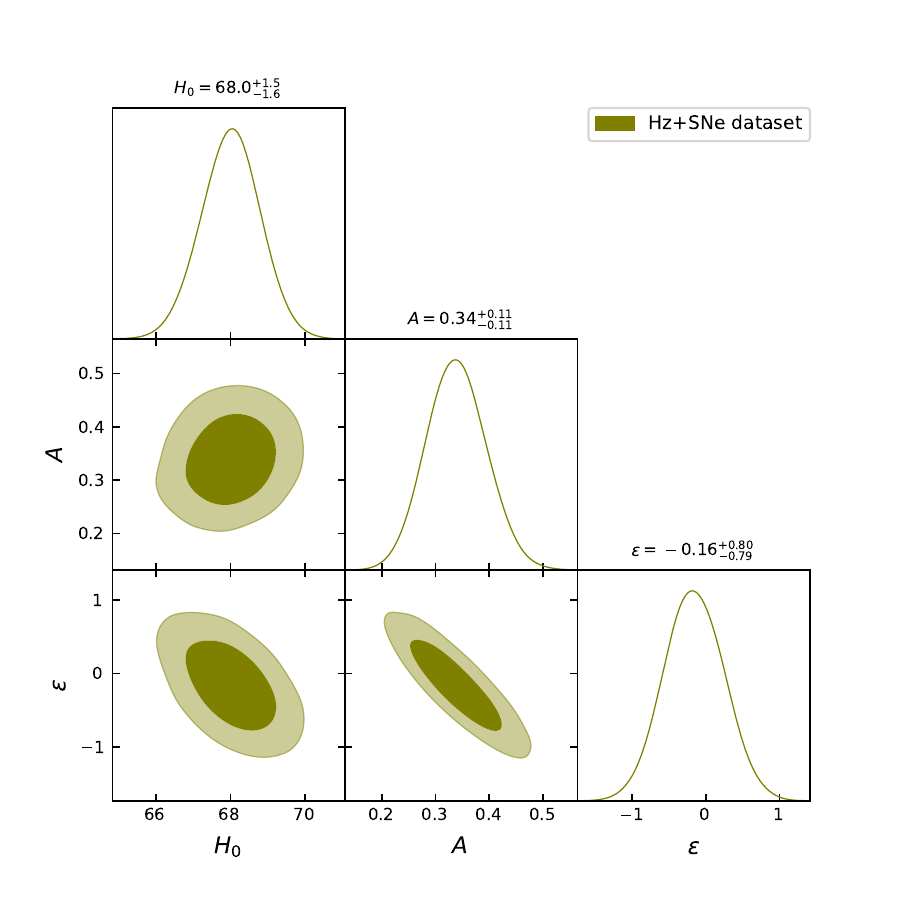}}
\caption{Likelihood contours for model parameters using the $Hz+SNe$ dataset: $1-\sigma$ and $2-\sigma$ confidence intervals depicted.}
\label{F_Hz+SNe}
\end{figure}

Fig. \ref{F_Hz+SNe} presents a contour plot illustrating the model parameters $H_{0}$, $A$, and $\epsilon$ based on the $Hz+SNe$ dataset. The contours depict the $1-\sigma$ and $2-\sigma$ confidence intervals. The obtained best-fit values for the model parameters are as follows: $H_{0}=68.0_{-1.6}^{+1.5}$ $km/s/Mpc$, $A=0.34_{-0.11}^{+0.11}$, and $\epsilon=-0.16_{-0.79}^{+0.80}$. This analysis provides an intermediate value between the measurement by Freedman et al. \cite{Freedman} and the value obtained from the Planck base-$\Lambda$CDM model \cite{planck_collaboration/2020}. 

\subsection{The deceleration parameter}

The deceleration parameter is defined as $q= -\ddot{a}a/\dot{a}^2 = -\ddot{a}/(H^2a)$. By considering Eq. \eqref{Hz}, we can derive the following expression:
\begin{equation}\label{23}
q(z)=\frac{A (1+z)^3-2 B-2 \epsilon  \log (1+z)+\epsilon }{2 \left(A (1+z)^3+B+\epsilon  \log (1+z)\right)}.
\end{equation}

\begin{figure}[h]
\includegraphics[scale=0.7]{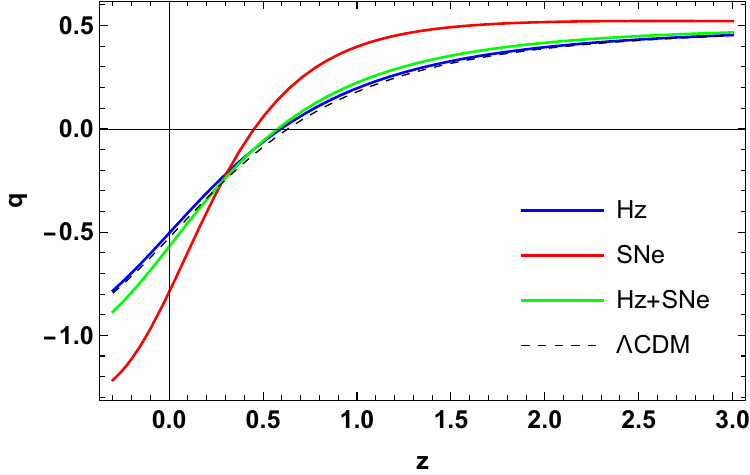}
\caption{Evolution of the deceleration parameter versus redshift $z$ for constrained model parameters using $Hz$, $SNe$, and $Hz+SNe$ datasets.}\label{F_q}
\end{figure}

The analysis of the deceleration parameter, as shown in Fig. \ref{F_q}, reveals a transition from a decelerated phase ($q>0$) to an accelerated phase ($q<0$) of the Universe's expansion at redshift $z_{tr}$, considering the constrained values of the model parameters. The present value of the deceleration parameter is found to be $q_0 = -0.51_{-0.27}^{+0.25}$ \cite{Garza} for the $Hz$ dataset, $q_0 = -0.79_{-0.01}^{+0.0}$ \cite{Cunha} for the $SNe$ dataset, and $q_0 =-0.57_{-0.56}^{+0.57}$ \cite{Camarena} for the $Hz+SNe$ dataset. Furthermore, the value of $z_{tr}$ of the model varies within the range of 0.4 to 1.0 as indicated by recent observations \cite{Mehrabi}. It is essential to emphasize that our model aligns well with the more widely accepted $\Lambda$CDM model when considering the $Hz$ and $Hz+SNe$ datasets. However, a notable deviation becomes apparent when comparing our model with the $SNe$ dataset. This indicates the sensitivity of our model's predictions to the specific dataset used for analysis.

\begin{figure}[h]
\includegraphics[scale=0.7]{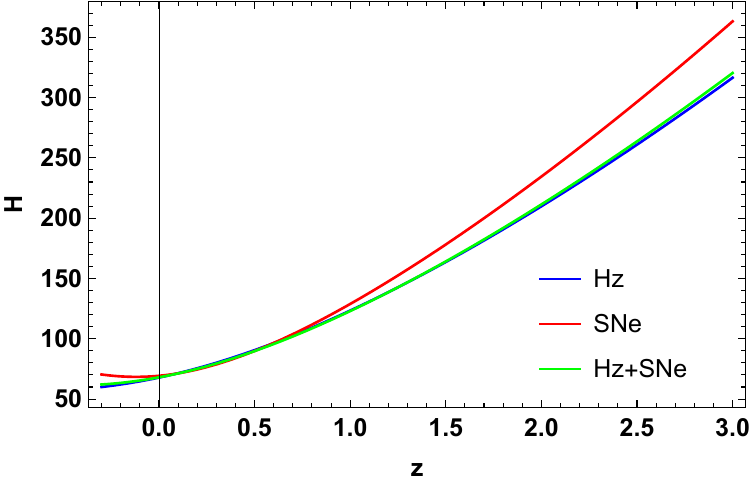}
\caption{Evolution of the Hubble parameter versus redshift $z$ for constrained model parameters using $Hz$, $SNe$, and $Hz+SNe$ datasets.}\label{F_H}
\end{figure}

Furthermore, as illustrated in Fig. \ref{F_H}, it is evident that the Hubble parameter demonstrates the expected positive behavior, in accordance with the model parameters derived from three distinct datasets: $Hz$, $SNe$, and the combined $Hz+SNe$.

\section{Cosmological $f(R,L_m)$ Models}\label{sec5}

In this discussion, we focus on cosmological models constructed using a logarithmic correction of the $\Lambda$CDM model. We aim to examine the stability of these models by testing the energy conditions. Specifically, we investigate two non-linear $f(R,L_m)$ models: $f(R,L_m)=\frac{R}{2}+L_m^\alpha$ and $f(R,L_m)=\frac{R}{2}+(1+\alpha R)L_m$, where $\alpha$ represents a free model parameter. These models are motivated by the generic $f(R,L_m)$ function, which takes the form of $f(R,L_m) = f_1(R) + f_2(R)G(L_m)$, representing arbitrary curvature-matter coupling \cite{Harko10}. Minimal and non-minimal coupling cases have gained significant attention from cosmologists in recent years, especially in the context of various modified gravity. The aforementioned generic $f(R, L_m)$ functions are particularly interesting as they encompass both minimal and non-minimal coupling scenarios. Through this exploration, we aim to deepen our understanding of the role played by minimal and non-minimal couplings in shaping the dynamics of DE in cosmological models.

\subsection{Model 1: $f(R,L_m)=\frac{R}{2}+L_m^\alpha$}

To study the dynamics of DE, we adopt a specific minimal $f(R, L_m)$ function \cite{Harko10,Jaybhaye}. This choice of minimal coupling case is motivated by the notable research conducted by \cite{Bose} within the framework of $f(R, T)$ gravity,
\begin{equation}
    f(R,L_m)=\frac{R}{2}+L_m^\alpha
\end{equation}
where $\alpha$ represents a free model parameter. Notably, when $\alpha=1$, we recover the standard Friedmann equations of GR. In the specific case of this $f(R, L_m)$ model, where $L_m=\rho$ \cite{Harko11}, the Friedmann equations \eqref{14} and \eqref{15} can be formulated as
\begin{equation}\label{F1}
3H^2=(2\alpha-1) \rho^{\alpha},
\end{equation}
\begin{equation}\label{F2}
2\dot{H}+3H^2=\left[(\alpha-1)\rho-\alpha p\right]\rho^{\alpha-1}.
\end{equation}

By using Eqs. (\ref{F1}) and (\ref{F2}), we can express the energy density $\rho$, pressure $p$, and the EoS parameter $\omega$ in terms of the Hubble parameter and its derivative with respect to cosmic time as
\begin{equation}
    \rho=\left(\frac{3H^2}{2 \alpha -1}\right)^{1/\alpha },
    \label{rho1}
\end{equation}
\begin{equation}
    p=-\frac{ \left(\frac{3H^2}{2 \alpha -1}\right)^{1/\alpha } \left(3 \alpha  H^2+(4 \alpha -2) \dot{H}\right)}{3\alpha  H^2},
    \label{p1}
\end{equation}
\begin{equation}
    \omega=\frac{p}{\rho}=-1+\frac{(2-4 \alpha ) \dot{H}}{3 \alpha  H^2}.
    \label{omega1}
\end{equation}

By substituting Eqs. (\ref{rho1}) and (\ref{p1}) into the energy conditions, we obtain the following expressions:
\begin{equation}
NEC\Longleftrightarrow -\frac{2\ 3^{\frac{1}{\alpha }-1} \dot{H} \left(\frac{H^2}{2 \alpha -1}\right)^{\frac{1}{\alpha }-1}}{\alpha }\geq 0,
\end{equation}%
\begin{equation}
    DEC\Longleftrightarrow \frac{2\ 3^{\frac{1}{\alpha }-1} \left(\frac{H^2}{2 \alpha -1}\right)^{1/\alpha } \left(3 \alpha  H^2+(2 \alpha -1) \dot{H}\right)}{\alpha  H^2}\geq 0,
\end{equation}
\begin{equation}
    SEC\Longleftrightarrow -\frac{2\ 3^{1/\alpha } \left(\frac{H^2}{2 \alpha -1}\right)^{1/\alpha } \left(\alpha  H^2+(2 \alpha -1) \dot{H}\right)}{\alpha  H^2}\geq 0,
\end{equation}

\begin{figure}[h]
\includegraphics[scale=0.7]{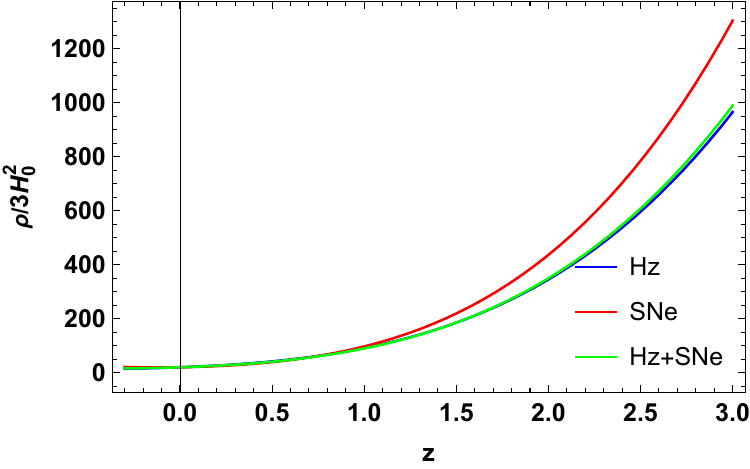}
\caption{Evolution of the energy density versus redshift $z$ for the model 1 ($\alpha=0.8$).}\label{F_rho1}
\end{figure}

\begin{figure}[h]
\includegraphics[scale=0.7]{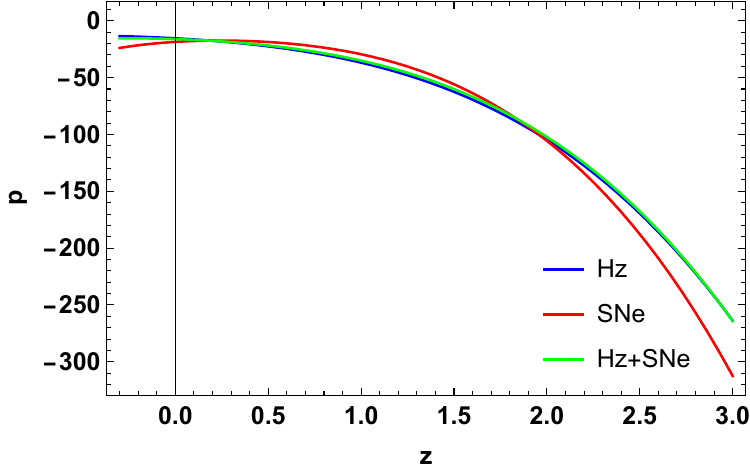}
\caption{Evolution of the pressure versus redshift $z$ for the model 1 ($\alpha=0.8$).}\label{F_p1}
\end{figure}

The plots below depict the behavior of the energy density $\rho$ and pressure $p$ as functions of redshift $z$ for the model parameters obtained from three different datasets: $Hz$, $SNe$, and the combined $Hz+SNe$. The equations (\ref{rho1}) and (\ref{rho2}) are utilized to calculate these quantities. The behavior of the $\rho$ and $p$ as a function of $z$ can be clearly observed in Figs. \ref{F_rho1} and \ref{F_p1}. The energy density shows an increasing function with redshift, indicating a positive value throughout the cosmic evolution. Conversely, the pressure exhibits negative values. These observations align with the expanding nature of the Universe, while the negative pressure signifies the presence of cosmic accelerated expansion. It is evident that the inclusion of the logarithmic correction of the $\Lambda$CDM model within the framework of the $f(R, L_m)$ gravity contributes to the phenomenon of DE.

The EoS parameter describes the relationship between the pressure $p$ and energy density $\rho$ within the Universe. It serves as a key factor in classifying the expansion behavior, distinguishing between decelerated and accelerated phases. The EoS parameter allows us to categorize different epochs based on specific values:
\begin{itemize}
    \item When the EoS parameter $\omega$ equals 1, it signifies a stiff fluid.
    \item For $\omega = \frac{1}{3}$, the model represents a radiation-dominated phase.
    \item A value of $\omega = 0$ corresponds to a matter-dominated phase.
    \item In the current accelerated phase of evolution, the range $-1/3 < \omega < -1$ indicates the quintessence phase. 
    \item The value $\omega = -1$ corresponds to the cosmological constant, which is the basis of the $\Lambda$CDM model. 
    \item Finally, for $\omega < -1$, the Universe enters the phantom era.
\end{itemize}

\begin{figure}[h]
\includegraphics[scale=0.7]{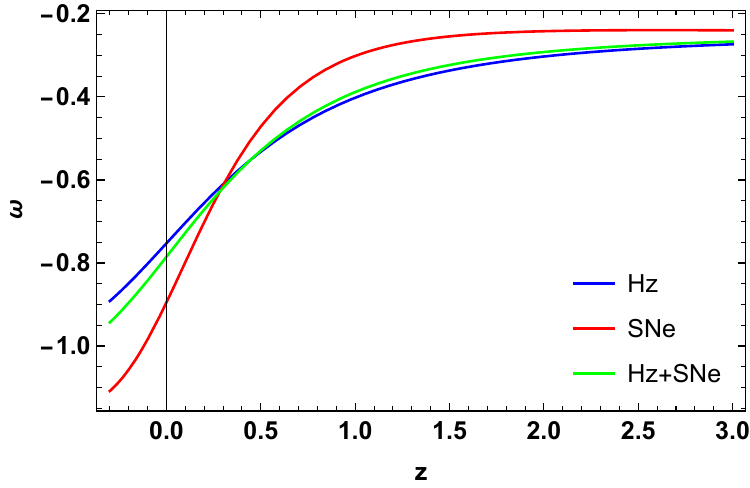}
\caption{Evolution of the EoS parameter versus redshift $z$ for the model 1 ($\alpha=0.8$).}\label{F_EoS1}
\end{figure}

Fig. \ref{F_EoS1} depicts the behavior of the EoS parameter with respect to $z$ for the model parameters obtained from three different datasets: $Hz$, $SNe$, and the combined $Hz+SNe$. According to Eq. (\ref{omega1}), when $z=-1$, the corresponding value of $\omega$ is $-1$. The graph presented in Fig. \ref{F_EoS1} illustrates that the model exhibits behavior consistent with quintessence DE. In addition, the present values of the EoS parameter for the $Hz$, $SNe$, and the combined $Hz+SNe$ are $\omega_0=-0.75_{-0.37}^{+0.38}$ \cite{Hernandez}, $\omega_0=-0.90_{-0.79}^{+0.79}$ \cite{Gong,Zhang}, and $\omega_0=-0.79_{-0.28}^{+0.29}$ \cite{Hernandez}, respectively.

\begin{figure}[h]
\includegraphics[scale=0.7]{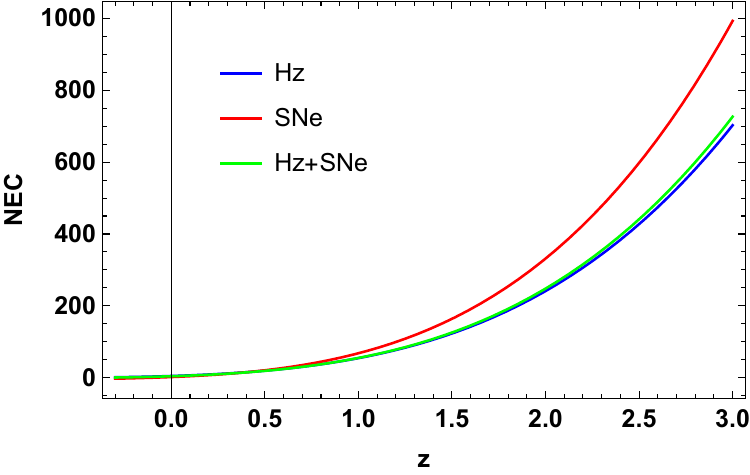}
\caption{Evolution of the NEC versus redshift $z$ for the model 1 ($\alpha=0.8$).}\label{F_NEC1}
\end{figure}

\begin{figure}[h]
\includegraphics[scale=0.7]{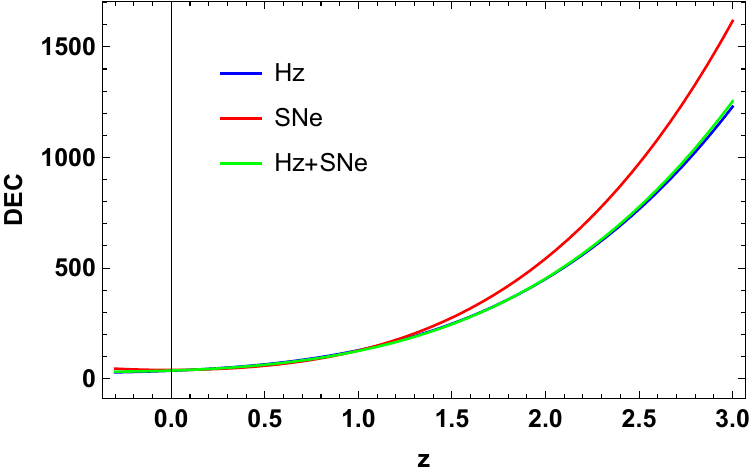}
\caption{Evolution of the DEC versus redshift $z$ for the model 1 ($\alpha=0.8$).}\label{F_DEC1}
\end{figure}

\begin{figure}[h]
\includegraphics[scale=0.7]{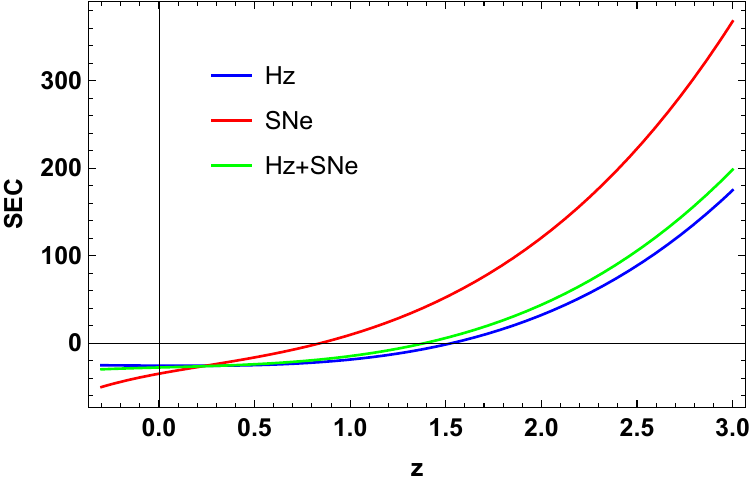}
\caption{Evolution of the SEC versus redshift $z$ for the model 1 ($\alpha=0.8$).}\label{F_SEC1}
\end{figure}

Below is the graph illustrating the behavior of the energy conditions. The primary objective of the energy conditions is to examine the expansion of the Universe. These conditions come in various forms, including the NEC, WEC, DEC, and SEC. The violation of the NEC indicates that none of the energy conditions mentioned are satisfied. The violation of the SEC is currently a topic of significant interest due to the observed accelerated expansion of the Universe \cite{Barcelo,Moraes}. According to cosmological scenarios, SEC must be violated during the inflationary expansion and at the present time \cite{Visser}. It is evident that for the model parameters obtained from three different datasets: $Hz$, $SNe$, and the combined $Hz+SNe$, both the NEC and DEC exhibit positive behavior. This indicates that both conditions are satisfied. Since the WEC is a combination of the NEC and the requirement of positive energy density, we can conclude that the WEC also holds for the model parameters obtained from three different datasets (see Figs. \ref{F_NEC1} and \ref{F_DEC1}). However, Fig. \ref{F_SEC1} shows that the SEC displays a transition from positive to negative behavior in the recent past. This violation of the SEC strongly supports the observed phenomenon of DE, indicating a transition from a decelerated phase to an accelerated phase.

\subsection{Model 2: $f(R,L_m)=\frac{R}{2}+(1+\alpha R)L_m$}

For the second model, we consider a specific non-minimal $f(R, L_m)$ function \cite{RL2,Kavya}. This choice of the non-minimal coupling case is motivated by various studies in the literature that have explored similar formulations,
\begin{equation}
    f(R,L_m)=\frac{R}{2}+(1+\alpha R)L_m
\end{equation}
where $\alpha$ represents a free model parameter. Notably, when $\alpha=0$, we recover the standard Friedmann equations of GR. In the specific case of this $f(R, L_m)$ model, where $L_m=\rho$ \cite{Harko11}, the Friedmann equations \eqref{14} and \eqref{15} can be formulated as
\begin{equation}\label{F2}
3 H^2 (6 \alpha  \rho -1)+12 \alpha  \dot{H} \rho +\rho =0,
\end{equation}
\begin{equation}\label{F3}
3 H^2 (-2 \alpha  \rho +4 \alpha  p+1)+\dot{H} (-2 \alpha  \rho +6 \alpha  p+2)+p=0.
\end{equation}

By using Eqs. (\ref{F2}) and (\ref{F3}), we can express the energy density $\rho$, pressure $p$, and the EoS parameter $\omega$ in terms of the Hubble parameter and its derivative with respect to cosmic time as
\begin{equation}
    \rho=\frac{3 H^2}{18 \alpha  H^2+12 \alpha  \dot{H}+1},
    \label{rho2}
\end{equation}
\begin{equation}
    p=-\frac{6 \alpha  \left(2 H^2+\dot{H}\right) \left(3 H^2+4 \dot{H}\right)+3 H^2+2 \dot{H}}{\left(18 \alpha  H^2+12 \alpha  \dot{H}+1\right) \left(6 \alpha  \left(2 H^2+\dot{H}\right)+1\right)},
    \label{p2}
\end{equation}
\begin{equation}
    \omega=\frac{p}{\rho}=-1+\frac{2 \dot{H}}{3 H^2}\left(\frac{1}{6 \alpha  \left(2 H^2+\dot{H}\right)+1}-2\right).
    \label{omega2}
\end{equation}

By substituting Eqs. (\ref{rho2}) and (\ref{p2}) into the energy conditions, we obtain the following expressions:
\begin{equation}
NEC\Longleftrightarrow -\frac{2 \dot{H} \left(12 \alpha  \left(2 H^2+\dot{H}\right)+1\right)}{\left(18 \alpha  H^2+12 \alpha  \dot{H}+1\right) \left(6 \alpha  \left(2 H^2+\dot{H}\right)+1\right)}\geq 0,
\end{equation}%
\begin{equation}
    DEC\Longleftrightarrow \frac{2 \left(6 \alpha  \left(2 H^2+\dot{H}\right) \left(3 H^2+2 \dot{H}\right)+3 H^2+\dot{H}\right)}{\left(18 \alpha  H^2+12 \alpha  \dot{H}+1\right) \left(6 \alpha  \left(2 H^2+\dot{H}\right)+1\right)}\geq 0,
\end{equation}
\begin{equation}
    SEC\Longleftrightarrow -\frac{6 \left(6 \alpha  \left(H^2+2 \dot{H}\right) \left(2 H^2+\dot{H}\right)+H^2+\dot{H}\right)}{\left(18 \alpha  H^2+12 \alpha  \dot{H}+1\right) \left(6 \alpha  \left(2 H^2+\dot{H}\right)+1\right)}\geq 0,
\end{equation}

\begin{figure}[h]
\includegraphics[scale=0.7]{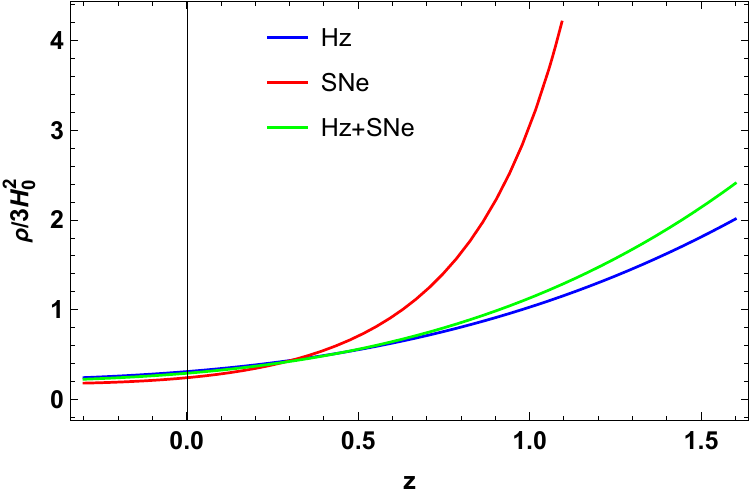}
\caption{Evolution of the energy density versus redshift $z$ for the model 2 ($\alpha=0.8$).}\label{F_rho2}
\end{figure}

\begin{figure}[h]
\includegraphics[scale=0.7]{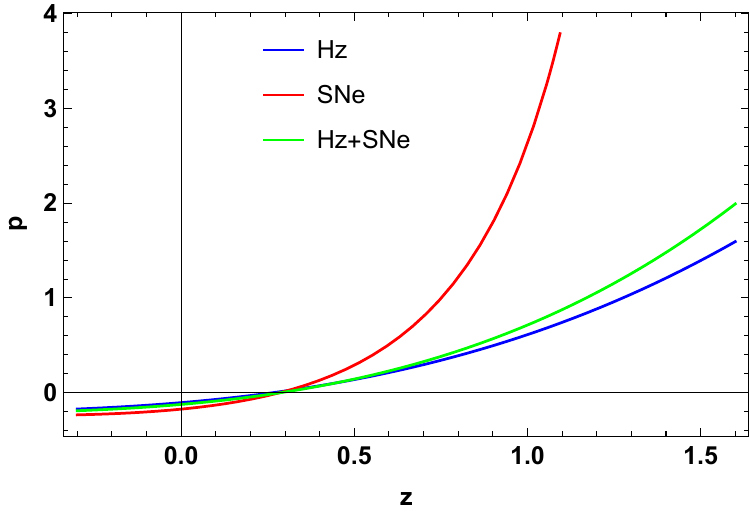}
\caption{Evolution of the pressure versus redshift $z$ for the model 2 ($\alpha=0.8$).}\label{F_p2}
\end{figure}

\begin{figure}[h]
\includegraphics[scale=0.7]{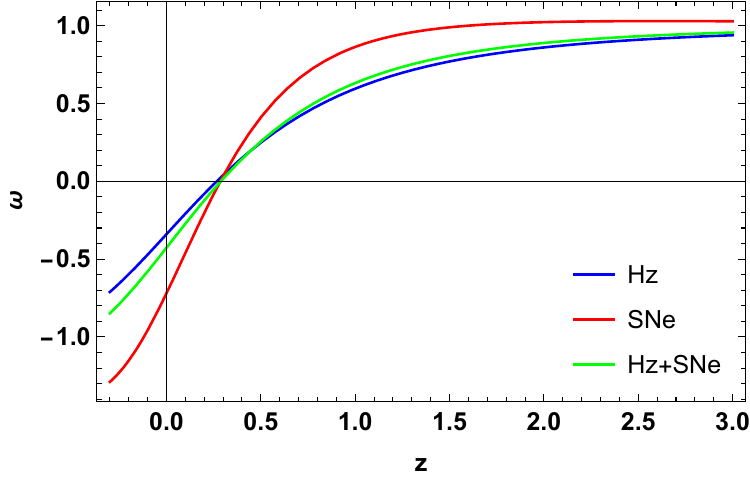}
\caption{Evolution of the EoS parameter versus redshift $z$ for the model 2 ($\alpha=0.8$).}\label{F_EoS2}
\end{figure}

In the case of the second model, as depicted in Fig. \ref{F_rho2}, the energy density exhibits a consistently positive behavior for all the constrained values of the model parameters, which aligns with our expectations. Furthermore, Fig. \ref{F_p2} illustrates that the pressure undergoes a transition from positive values in the past to negative values in both the present and future. In Fig. \ref{F_EoS2}, the EoS parameter is plotted as a function of redshift z, using the same constrained values of the model parameters. The figure illustrates the transition from negative to positive values over the course of cosmic evolution. This transition indicates an earlier decelerating phase of the Universe characterized by positive pressure, suitable for structure formation. In the present accelerating stage of evolution, the EoS parameter exhibits negative pressure. Furthermore, it is evident that the present values of the EoS parameter ($\omega_0=-0.34_{-0.31}^{+0.33}$, $\omega_0=-0.72_{-0.65}^{+0.67}$, and $\omega_0=-0.43_{-0.11}^{+0.10}$), in this case, satisfy the condition for the acceleration phase, i.e., $\omega_{0}<-\frac{1}{3}$ \cite{Koussour1,Koussour2}.

\begin{figure}[h]
\includegraphics[scale=0.7]{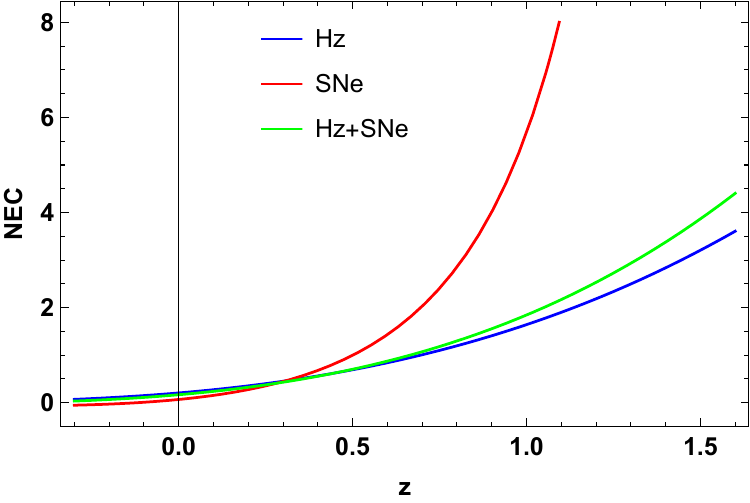}
\caption{Evolution of the NEC versus redshift $z$ for the model 2 ($\alpha=0.8$).}\label{F_NEC2}
\end{figure}

\begin{figure}[h]
\includegraphics[scale=0.7]{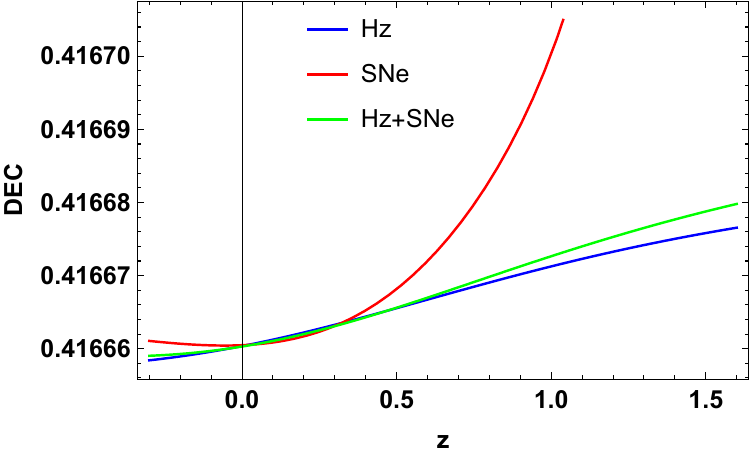}
\caption{Evolution of the DEC versus redshift $z$ for the model 2 ($\alpha=0.8$).}\label{F_DEC2}
\end{figure}

\begin{figure}[h]
\includegraphics[scale=0.7]{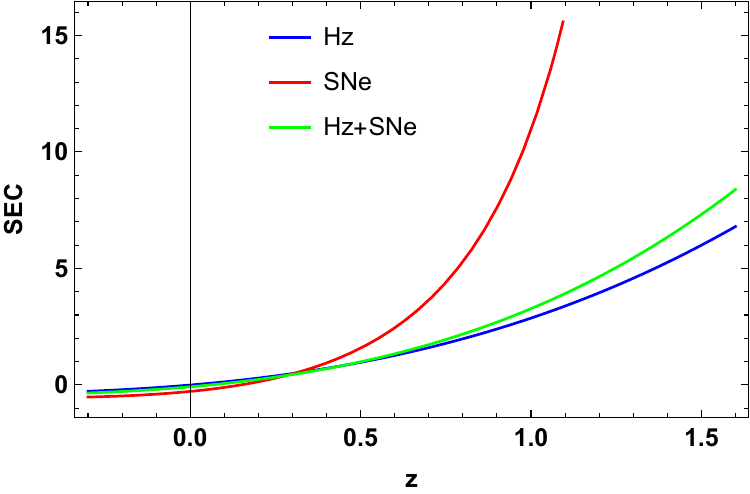}
\caption{Evolution of the SEC versus redshift $z$ for the model 2 ($\alpha=0.8$).}\label{F_SEC2}
\end{figure}

In the case of the second model, a similar analysis of the energy conditions was conducted (see Figs. \ref{F_NEC2}, \ref{F_DEC2}, and \ref{F_SEC2}). For the model parameters obtained from the three different datasets ($Hz$, $SNe$, and the combined $Hz+SNe$), both the NEC and DEC exhibit positive behavior, indicating their satisfaction. The WEC also holds for these model parameters. However, the SEC shows a transition from positive to negative behavior in the recent past, violating the condition set by the SEC. This violation of the SEC further supports the observed phenomenon of DE.

\section{Conclusion}\label{sec8}

Currently, both dynamical DE models and alternative theories of gravity are widely explored to explain the observed accelerated expansion of the Universe. Among them, $f(R,T)$ and $f(R,L_m)$ theories have gained significant attention in the past. In this study, we have focused on spatially homogeneous and isotropic FLRW cosmological models, considering the presence of a perfect fluid within the framework of $f(R,L_m)$ theory. We have investigated two non-linear functional forms of this theory to understand their implications on cosmic evolution. Specifically, $f(R,L_m)=\frac{R}{2}+L_m^\alpha$ (model 1) and $f(R,L_m)=\frac{R}{2}+(1+\alpha R)L_m$ (model 2), where $\alpha$ represents a free model parameter. Further, we have employed a parametrization form of the Hubble parameter in terms of redshift $z$ as $H(z)=H_0 \left[A(1+z)^3+B+\epsilon \log(1+z)\right]^\frac{1}{2}$, to obtain the cosmological solution. Subsequently, we have incorporated this solution into the Friedmann equations of $f(R,L_m)$ gravity.

To assess the viability of this model and its potential deviations from the $\Lambda$CDM framework, we conducted a Bayesian analysis to derive constraints on the model parameters using $Hz$ measurements (31 points) and $SNe$ dataset (1048 points). The best-fit values obtained for the model parameters are $H_{0}=67.8_{-1.8}^{+1.8}$ $km/s/Mpc$, $A=0.33_{-0.13}^{+0.14}$, and $\epsilon=0.0_{-1.1}^{+1.1}$ for the $Hz$ dataset, $H_{0}=69.3_{-4.0}^{+4.1}$ $km/s/Mpc$, $A=0.44_{-0.31}^{+0.32}$, and $\epsilon=-0.9_{-2.2}^{+2.2}$ for the $SNe$ dataset, and $H_{0}=68.0_{-1.6}^{+1.5}$ $km/s/Mpc$, $A=0.34_{-0.11}^{+0.11}$, and $\epsilon=-0.16_{-0.79}^{+0.80}$ for the $Hz+SNe$ dataset. In addition, we examined the evolution of key cosmological quantities, including the deceleration parameter, energy density, pressure, EoS parameter, and energy conditions for the model parameters obtained from three different datasets: $Hz$, $SNe$, and the combined $Hz+SNe$. The analysis reveals that the deceleration parameter undergoes a smooth transition from a decelerated phase to an accelerated phase of expansion. Additionally, as depicted in Figs. \ref{F_rho1} and \ref{F_rho2}, the energy density for both models decreases as the Universe continues to expand in the distant future, consistently maintaining positive values throughout the cosmic evolution. Conversely, the pressure, as illustrated in \ref{F_p1} and \ref{F_p2}, exhibits negative values in the present and future. The EoS parameter in Figs. \ref{F_EoS1} and \ref{F_EoS2} display negative behavior, indicating an accelerating Universe and the presence of quintessence DE. Notably, the present values of the deceleration and EoS parameters align well with the most recent observations of cosmological parameters and are consistent with previous studies in the field \cite{Hernandez,Gong,Zhang}. In addition, the findings presented in this study are consistent with various models of DE based on empirical observations, including models such as $f(Q)$ and $f(Q,T)$ gravity. The behavior of quintessence, a fundamental component in these models, is evident in the results obtained. These observations further underscore the compatibility of the $f(R,L_m)$ gravity theories with a wide range of DE models and their potential to provide a robust framework for describing the cosmic evolution \cite{Koussour1,Koussour2,Koussour3,Koussour4,Koussour5}.

In the final analysis, the energy conditions were examined for both models to assess the validity of the obtained solution. It was observed that all energy conditions, except for the SEC, exhibited positive behavior, as depicted in Figs. \ref{F_NEC1}, \ref{F_DEC1}, \ref{F_NEC2}, and \ref{F_DEC2}. However, the violation of the SEC, as shown in Figs. \ref{F_SEC1} and \ref{F_SEC2}, strongly support the accelerating nature of cosmic expansion and signify a transition from a decelerated phase to an accelerated era. This further corroborates the evidence for the presence of the DE phenomenon driving the accelerated expansion of the Universe in the framework of $f(R,L_m)$ gravity.

\section*{Data Availability Statement}
There are no new data associated with this article.


\end{document}